# The New War Correspondents:
# The Rise of Civic Media Curation in Urban Warfare

Andrés Monroy-Hernández   danah boyd   Emre Kiciman
Munmun De Choudhury   Scott Counts
One Microsoft Way, Redmond WA 98052
{andresmh, dmb, emrek, munmund, counts}@microsoft.com

**ABSTRACT**

In this paper we examine the information sharing practices of people living in cities amid armed conflict. We describe the volume and frequency of microblogging activity on Twitter from four cities afflicted by the Mexican Drug War, showing how citizens use social media to alert one another and to comment on the violence that plagues their communities. We then investigate the emergence of civic media "curators," individuals who act as "war correspondents" by aggregating and disseminating information to large numbers of people on social media. We conclude by outlining the implications of our observations for the design of civic media systems in wartime.

**Author Keywords**

social media; social computing; civic media; crisis informatics; war; crowdsourcing; curation; news; microblogging; Twitter; Latin America; Mexico

**ACM Classification Keywords**

K.4.3 [Computers and Society]: Organizational Impacts–Computer-supported cooperative work. H.5.3 Groups & Organization Interfaces—collaborative computing, computer-supported cooperative work; K.4.2 Social Issues

**INTRODUCTION**

Since 2006, Mexico has been in the middle of a drug war as cartels engage in urban combat throughout the country. As of 2011, this ongoing, large-scale conflict has taken the lives of more than 60,000 people [10,44], and has displaced more than 230,000 people [43].

Frequent clashes among drug cartels, the military, and police forces take place in urban areas, inducing panic, cases of post-traumatic stress disorder [26], and, in some cases, taking the lives of innocent bystanders.

As drug war violence spreads and traditional news media weakens, frustrated citizens turn to the increasingly pervasive social media for "information and survival" [9]. Twitter in particular has become one of the principal sources of citizen-driven alerts in several Mexican cities; people often report, confirm, comment on, and disseminate information and alerts about the violence, typically as it unfolds. For example, the following Twitter message reports the time and location of blasts, along with a list of hashtags or keywords that both label and enable discovery of shared information resources:

> "There are reports of blasts on Venustiano Carranza Avenue #Shooting #RiskMty #MtyFollow."[1]

In the first part of this paper we examine a large corpus of messages like the one above, using observational and quantitative analyses, to understand how people living amid urban warfare in Mexico use microblogging to alert one another about frequent and often expected violent clashes. In the second part of this paper we examine, through quantitative and conversational techniques, the role of a small group of individuals who have a large audience and contribute a sizable amount of content related to acute events in specific cities.

This paper refines and extends the existing literature on crisis and emergency informatics by observing the longitudinal evolution of social media practices among people living in a setting where acute events cannot be temporally reduced to "uprisings" or "disasters," and where institutions are weakened.

**PREVIOUS WORK**

The phenomenon of people using social media during crisis is not new. Researchers have examined how people use social media in crises like floods [6,35] earthquakes [36], terrorist attacks [11], school shootings [29], and political uprisings [1,24]. Although all crises have attributes in common, those that occur as a result of warfare are qualitatively different: they are a part of everyday life, turning otherwise extraordinary events into ordinary ones.

**Social Media in Wartime**

---

---

[1] Texts in Spanish are translated to English by the authors for exemplary purposes. To the extent possible, we maintain the style of the content translated from Spanish.

Research on social media use during the Iraqi war, for example, has described how blogs helped Iraqi citizens report events as they unfolded [2], maintain a routine [25], and enabled the discussion of topics that were socially unacceptable to discuss face-to-face. They even changed socialization practices, such as finding marriage partners [29]. Drawing on Powell's model of crises [30], Seeman and Mark [34] describe how people use social media differently depending on where they are in the crisis; they point out that war presents a special case for crisis informatics because people are experiencing several stages concurrently and, thus, using social media to serve different needs.

### Government's Role in Emergencies

Another angle examined in the literature is the role of government officials. Typically, during crisis, governments attempt to play a helpful and central role in trying to restore normalcy, partly by brokering information. For example, in the United States, Public Information Officers (PIOs) are government officials "charged with coordinating communication activities and performing as spokespersons" during crises [22]. Researchers have found that with the advent of social media, the role of the PIO has changed: PIOs now can interact directly with the public on social media but also have to deal with fast-spreading misinformation on these new platforms [20].

Many government agencies leverage technology to inform the public, but what is provided is often incomplete or insufficient [18]. More troubling, some governments – such as China during the SARS epidemic [31] – use technology to purposefully misinform the public. In response to this, social media has also provided a venue to present a counter narrative to government-controlled media, as it was in the recent Egyptian revolution [1].

Yet, even with the increasing power of social media and other information technologies, in countries like the United States, PIOs and mainstream media are still the authoritative sources of information. However, in war zones, PIOs are weakened by conflict and are unable or unwilling to carry out their role. In such situations, government information is often unreliable [34]. As we will describe, this is the context in Mexico as the Drug War persists.

### Social Media Curation

Twitter, Facebook, YouTube, and similar sites and services represent a genre of computer-mediated communication tools colloquially referred to as social media [12]. Social media enables participants to contribute content and share others' content. The resulting ecosystem of "user-generated content" is often contrasted with content produced by experts and pundits debating the merits of crowdsourced material. Yet, when people engage with this content, they help create a participatory culture surrounding what Henry Jenkins at al. [21] call "spreadable media." People take on numerous roles in this ecology, helping produce and disseminate content while also providing commentary and critique.

Often, for want of a better word, those who take an active role in spreading information through social media – with or without commentary – are colloquially identified as "social media curators." This term is fraught, given the precise roles that professional curators, librarians, archivists, and art collectors tend to play. Like other terms corrupted by social media (e.g., friend, social network), the concept of a social media curator metaphorically draws on, but is not equivalent to, the more precise term.

Social media curators seek to spread information to new audiences by selectively identifying and sharing content coming from the broader stream. These curators develop reputations with their audiences based on the perceived value of the information that they spread. Some curators simply pass on information posted by others, while other curators add commentary or insert their own interpretations or updates. While we recognize that this term is loaded, we use it throughout this paper to refer to those actors who position themselves as intermediaries in the flow of spreadable media.

Research on crisis informatics has started to examine decentralized efforts at news curation, such as those using systems like Ushahidi and wikis [23], where large numbers of somewhat homogenous participants collaborate to archive information about crises like the 9-11 terrorist attacks. As Sarcevic et al. [33] note, crises are inherently decentralized events, but social media and other networked technologies can play a crucial coordination role. Researchers have noted that during crises, mainstream media often invites citizens to share their reports via social media [20].

### Weakening of Institutions in the Mexican Drug War

Like other armed conflicts, the Mexican Drug War is also a conflict over the control of information. Local news media organizations and governments have been forced into self-censorship and, as some claim, into collaborating directly with criminal groups.

Local news media organizations throughout the country have experienced extreme pressure not to report on the violence that engulfs the localities they serve. For example, a newspaper in Saltillo declared in an editorial [40] that due to the "threats on their editorial staff," they were "obligated, on occasion, to leave out information." Similarly, after several of their journalists were killed by drug gangs, a newspaper in Ciudad Juarez directed its front page headline to them, asking: "What do you want from us?" [39]. More recently, a newspaper office in Veracruz was burned down by criminal gangs [46]. Not surprisingly, a report [45] by the Committee to Protect Journalists ranked Mexico as the third most dangerous country for journalists. The result, according the foreign press [5], is a "near-complete news blackout" imposed by criminal organizations that "via daily telephone calls, e-mails and news releases" decide "what can and cannot be printed or aired."

Local governments, likewise, fail in public communication. For example, the foreign press reported how the mayor of a northern city "mysteriously disappears for days and refuses to discuss drug violence" [5]. Citizens' comments on social media and the comment sections of newspapers are filled with speculations on the reasons for this silence: fear of reprisals, collaboration with drug cartels, obliviousness, or an attempt to maintain an image of having everything under control.

Moreover, on those occasions when traditional news media or the government does report violence, the information is often published several hours later or even the day after the events take place. Hence citizens find these institutions ineffective at alerting them and helping them avoid violent clashes.

**Social Media Usage in Mexico**
Internet and social media use in Mexico have increased significantly in recent years, but neither is by any means universal. In order to situate our analysis of how social media is used to spread information, it is important to put social media usage in context. According to a report by the Mexican Internet Association [42], the number of people in the country with access to the Internet increased from 17.2% in 2000 to 34.9% in 2010. The report estimates that most of the 34 million people with Internet access are young: 37% are under 18, and 40% are between 18 and 34. The report also estimates that social media sites are used by 61% of people with Internet access and that of those, 39% use Facebook, 38% use YouTube, and 20% use Twitter. 53% of Twitter users reported using it at least once a day, from their home (39%), work (16%), and on mobile devices (18%). While Internet access in Mexico is neither universal nor uniformly distributed, social media does serve as an information channel that can reach a sizable percentage of the population. Furthermore, if Mexican citizens are anything like their American counterparts [32] younger citizens are more likely to turn to the Internet first for local news while older citizens are more likely to explore traditional media outlets first.

**SOCIAL MEDIA USAGE IN THE MEXICAN DRUG WAR**
In this section we describe how citizens are using social media as a resource for alerts of local violence in four cities embattled by the Mexican Drug War. There has been considerable attention in the media about Mexican citizens using Twitter to report violent events in their communities [7,9,14,27,38]. However, little is known concretely about the actual volume, content, or people involved. In this section, we provide a description and interpretation of the participation patterns in these cities over the course of sixteen months as manifested on the social media service Twitter. The choice of these quantitative metrics derives from aspects typically found to be of value in the crisis literature referenced in the previous section.

**Methodology**
We used a mix of observational, quantitative, and conversational techniques to understand how citizens are using Twitter to share information about the Mexican Drug War.

First, to get the lay of the land, we began our project by engaging in a variety of observational exercises involving the following processes:

a) *Searching and browsing Twitter*: Since early 2010, we have been informally monitoring Twitter for references of the violence occurring in Mexico. A key task of this step was to identify the types of hashtags that were being used by individuals in Mexico for violence reporting.
b) *Identifying co-occurring hashtags*: Once we identified a hashtag being used in tweets about violence, we started to search for others included in the same message. This was meant to improve the coverage of Twitter posts that are likely to contain information about events related to the Drug War.
c) *Monitoring trends*: Since mid-2011, we have been monitoring Twitter Trending Topics[2] in Mexico, which often are related to violence, through qualitative observations. When a trend related to violence would appear we would collect relevant data for further analysis.
d) *Reading news reports*: Once Twitter became known as a microblogging service where people were alerting each other, the news media started to report about this phenomenon and even use Twitter as the source for their articles related to the violence.
e) *Following high profile users*: Occasionally journalists, news media organizations, government officials, and public figures would amplify information from social media; this helped us to identify additional hashtags and gain a better sense of the ecosystem.

In collecting qualitative, observational data as a form of digital fieldwork, our goal was primarily to get an embedded sense of how digital media was being employed by citizens and understood by the media. This process benefited from our familiarity with the Spanish language, Mexico, Mexican media, and the Drug War.

| City | Population | Hashtag | Tweets |
|---|---|---|---|
| **Monterrey** | 4,000,000 | #mtyfollow | 211,278 |
| **Reynosa** | 600,000 | #reynosafolllow | 155,786 |
| **Saltillo** | 820,000 | #saltillo | 153,879 |
| **Veracruz** | 702,000 | #verfollow | 87,801 |

Table 1: Cities in the data set

Through our observations on Twitter, we identified a set of four cities central to discussions about the Mexican Drug

---
[2] Twitter's proprietary mechanism to demonstrate what type of topics are of high interest in a geographic region at a given point in time.

War: Monterrey, Reynosa, Saltillo, and Veracruz. For each of these cities, we identified a corresponding hashtag for analysis (see Table 1). We report their population size as a coarse comparative metric.

For each city, we identified the oldest and most commonly used hashtag in connection with violence reports for that city. Using the Twitter "firehose," available for use to our company through a contract with Twitter, we collected all public tweets containing the hashtags in Table 1 over the course of 16 months, from August 2010 to November 2011. This resulted in a corpus of 609,744 tweets for the four hashtags listed on Table 1.

**Results**

First, we found that, except for Reynosa, one third or more of the tweets (29.9% to 40%) were *retweets,* while a fifth were *mentions*, that is, tweets that included the name of another user (see Table 3). The high proportion of retweets over posting new content suggests that for those three Mexican cities, spreading information is the preferred mode of contributing to the discussion.

Second, we found that the number of daily tweets goes up and down constantly, spiking when violence erupts and decreasing when the city is calm (see Figure 1). The biggest spike in Monterrey (7,027 tweets) occurred on August 25, 2011, the day of an attack at a casino where 53 people were killed [3]. During this event people shared images of the attack and later on the names of missing people that family members presumed might have been at the casino. For Veracruz, the biggest spike (4,699 tweets) also happened on August 25, 2011, during a seemingly unrelated event, when rumors of school children being kidnapped spread on social media [29]. The biggest spike in Reynosa (2,745 tweets) happened on November 6, 2010 after intense clashes erupted due to the capture of a prominent drug lord [27]. For Saltillo, the biggest spike (4,637 tweets) occurred on October 12, 2011, also linked to clashes caused by the capture of a drug cartel leader [44].

| City | Daily Median | Mentions | Retweets | Users |
|---|---|---|---|---|
| **Monterrey** | 340 | 20.5% | 40.1% | 27,170 |
| **Reynosa** | 249 | 24.8% | 15.9% | 9,043 |
| **Saltillo** | 219 | 19.9% | 29.9% | 16,347 |
| **Veracruz** | 494 | 22.5% | 35.3% | 12,522 |

Table 3: Aggregate results of tweets per city

Third, doing a simple word (unigram) count and aggregation of the most common words by type (see Table 2), we found that the most common words are those used to refer to *places* in the city (such as the name of streets and neighborhoods), followed by words associated with *shootings*, followed by the verb "to report," followed by generic people references. Interestingly, the hashtag for a different city (Reynosa) is also one of the most common words in Monterrey.

Fourth, we found 65,082 unique posters across the four cities, who, on average, posted 9.4 tweets each. More specifically, those who posted new reports did so at a rate of 10.8 per person, those who spread tweets did so at a rate of 4.1 retweets per person, and those who referenced others did so at a rate of 5.5 per person. A few users contributed most of the content. For example, in the hashtag associated with Monterrey, half the people (52.8% or 14,898) tweeted only once, while a tiny fraction (0.03% or 9 people) tweeted more than a thousand times each. In the next section we analyze this activity difference in more detail.

| Type of Words | Words (Spanish) |
|---|---|
| places | zona, san, sur, altura, garza, col., av. |
| shootings | #balacera, balacera, balazos, detonaciones |
| report | reportan |
| people | gente, alguien |
| city names | #reynosafollow |

Table 2: Most common word types in Monterrey

**Discussion**

The popularity (i.e., high proportion) of retweets over posting new content suggests that for those Mexican cities, spreading information is the preferred mode of contributing to the discussion over interactions.

This would also explain the common use of Twitter for other purposes, including the search for blood donations, and even commercial purposes, which is often frowned upon by the community, as exemplified in this tweet:, "Sirs, I have created #mtyfollou so you can write your f*** stupidities there and stop misusing #mtyfollow."

As for Reynosa, the city with a bigger proportion of mentions, we find that for many people, Twitter is not only about alerts but also about interacting with others and creating bonds among those participating in the hashtags. For example, Reynosa tweets in the mornings are often about saying "good morning" to the members of a hashtag.

Our findings suggest that the adoption of Twitter did not happen simultaneously across the four cities. For example, Veracruz adopted this medium much later (see Figure 1). A likely explanation is that the practice spread as the violence spread. For example, we identified the first instance of the #verfollow where one person tells another of the possibility of using the hashtag #verfollow to find information about Veracruz: "you can find information about the state in verfollow or alertavera, messages should be short." Shortly after, another person sent a tweet to an influen-tial/authoritative Twitter user informing that user of the new hashtag: "there is a new follow for Veracruz: #verfollow." It is interesting to note that the poster used the word follow as a noun.

Our data also suggests that, in the cities we examined, there is widespread adoption of Twitter. If we assume that each account represents a unique person, our data suggests that 1.48% of people living in the cities analyzed posted something on Twitter about the Mexican Drug War. Assuming these cities have similar Internet penetration as the rest of Mexico (34.9% of the population, as cited above), an average of 4.2% of the online population has posted something about the Drug War on Twitter. These estimates have many limitations because they assume that: 1) people tweeting with those hashtags live in the city associated with the hashtag, 2) Internet penetration is homogeneous, and 3) people have only one account. Conversely, not every person who uses the hashtags was analyzed. Additionally, Twitter's estimate that 40% of their active users sign in to "listen" [37], might also suggest that the percentage of the population reading tweets about the Drug War is higher than those posting about them. Although it is almost impossible to tell precisely how many people are engaged in discussions surrounding the Drug War, it is not negligible.

## THE EMERGENCE OF CURATORS

In the previous section, we saw how the volume of activity varied across different participants. In this section, we focus on those people who not only have significant contributions (i.e., tweet count), but also have a sizable audience (i.e., follower count).

### Methodology

By analyzing each user's follower and tweet counts, we found distinct participation patterns. To better understand the distribution of participants, we present a scatter plot visualization depicting the relationship between tweet frequency and follower count (see Figure 2). First, a small number of highly-followed users – media organizations, celebrities, etc. – had contributed just a handful of tweets. For example, the account for "CNN en Español" has more than a million followers but had tweeted only once with the Monterrey hashtag. Meanwhile, a small number of users have a lot of followers and have contributed a lot of tweets (located towards the top right corner of the plots). We decided to examine those users.

In order to get a better understanding of curators' motivations and practices, we began reaching out to individuals who took on high profile curatorial roles on Twitter. As we found in our quantitative analysis, these curators often played a central role in helping spread information but they rarely provided their names on their profiles. As we learned through talking with them, the choice to protect their real identity is a conscious one.

We put together a list of the most active users in just the city of Monterrey by selecting those individually responsible for 1% or more of the city's tweets. This resulted in a list of ten users: one was responsible for posting 3% of all tweets, two were responsible for 2%, and the remaining seven were each responsible for 1% of the posts.

Initially, we had hoped to interview the various curators, but getting responses proved difficult. Although one coauthor approached a dozen of these users in Spanish as a researcher and Mexican national, they reasonably questioned the motives behind the project. Those who did respond (4 in total) often preferred not to talk outside of Twitter, so we began conversing with different curators in 140 character chunks, either publicly or through direct message. In some cases, we were able to transition our conversations to email or get them to respond to longer questions that we put up on a website. One curator agreed to be interviewed using a voice call on Skype.

It was clear during our recruitment and interviews that trust was a major issue. None of those willing to speak were willing to identify themselves and regularly refused to answer questions we asked. Still, the insights that we received

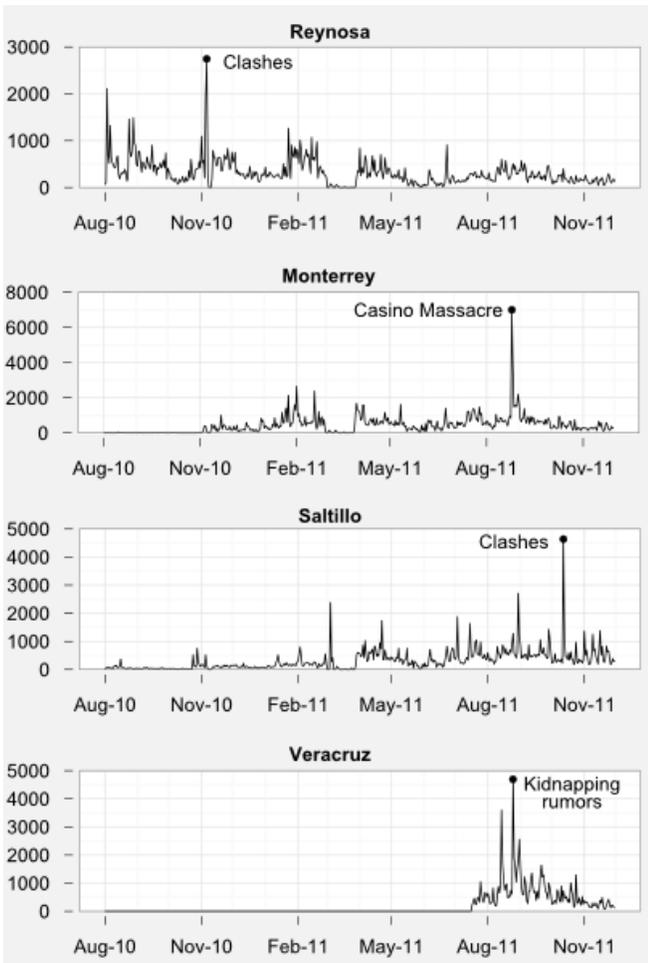

**Figure 1: Tweets per day for each of the cities' hashtags**

proved valuable, shedding light on the found data collected through computational means.

### Results

A closer look at the activity of those users with a high tweet *and* follower count over the course of the months we examined, shows that these people have taken the role of both aggregating and disseminating information to a large number of people in the city who follow them and who even send reports to them. These people, whom we call "curators," play a role akin to the PIOs in the United States, but through completely citizen-driven efforts.

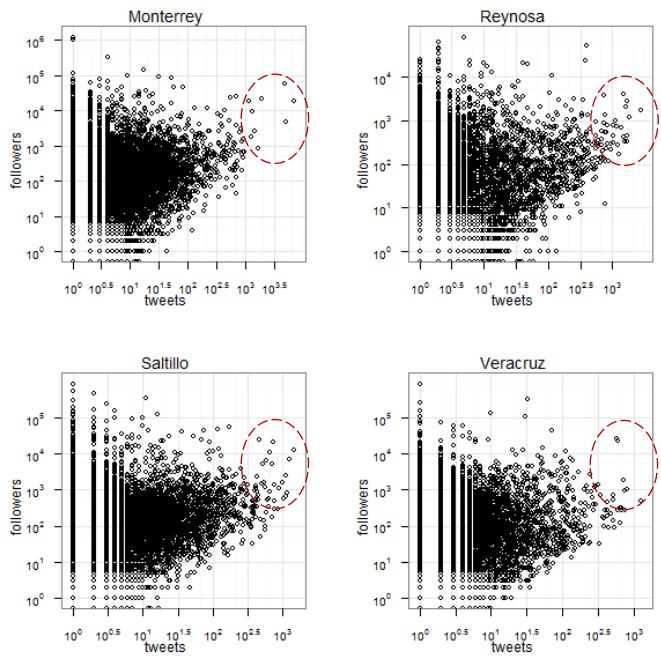

Figure 2: Followers and tweets for all users in each city. Curators region circled.

Some of those who posted frequently had few followers, while others had many followers (median: 9,305). Likewise, some accounts were regularly mentioned by others while others received few mentions (median: 303). The four accounts with more followers and mentions than the median all engaged in information curation, that is, receiving, responding to, and retweeting dozens of tweets from other accounts. Examining the dynamics of these accounts sheds light on the kinds of practices that are emerging in this ecosystem.

It is important to note that, besides these information curators, the other actors with a significant number of followers in the city are government officials (for example, the governor, the city police, and so on), news media (for example, newspapers and TV), and journalists. Together, four curators in Monterrey have 115,678 followers, almost three times the followers of the governor @RodrigoMedina (40,822) and almost as many as the most popular news media organization @Telediario (139,919).

### Motivations: Altruism and Attention

Three of the accounts were created in 2009, the year when the violence in the city first spiked. Some joined mainly for personal reasons, or to follow celebrities. For example, "Angela," a woman in her early twenties, joined Twitter because a friend encouraged her to join, while "Claudia" started using Twitter because of the chance of interacting with celebrities: "I heard on the radio about how artists and celebrities would interact with their fans, so I got curious about it." All three of these curators identify themselves as women who want to do something to address the violent situation of their city. For example, "Claudia" described herself as "Praying for god to bless Monterrey," while "Alex" claims to be the "founder" of the concept of "city surveillance" and part of a "citizen network to protect and provide tips about civic safety, to avoid becoming victims of crime."

When asked, all of them mentioned feeling abandoned by their government and the media. "Angela" for example, when asked for her opinions on mainstream media, answered:

"…they forgot they have an obligation with the people… they started hiding information. Then, society started demanding information. This is when social networks took over."

When asked about their motivations, two of these curators cited *altruistic* reasons. "Angela" said "I always make clear that tweeting is an altruistic community service," before adding "it's like if I was a *news correspondent* on social networks of the war we are living." "Claudia" said that people tell her she is "like their angel for looking after them" to which she added: "that is one of the strongest motivations for me to continue tweeting daily!" Both of them mentioned being on top of their Twitter account throughout the day; "Angela" reported spending about 15 hours a day while "Claudia" just said "many hours."

The fourth account, *CIC* (which stands for Center for Civic Integration), emerged in 2011 and describes itself as a "network of trust, 100% driven by citizens that provides clear, reliable and authentic spaces for participation and to strengthen citizenship and quality of life." Unlike the other more organic, pseudonymous, and sometimes controversial curators, this account presents itself as a civic organization with a public presence in the media, staff, and offices. CIC also receives reports through e-mail, Facebook, and phone, and has staff that continuously checks these various communication channels and updates an online map mash-up. Although the organization was born out of citizens' frustration with the violence, CIC has expanded its reach to cover all sorts of urban issues such as problems with street signs and even psychological counseling for those suffering from the effects of the violence. Unlike the other three curators, this account maintains a more formal, serious, business-like presence online, while the other curators take a personable approach that implies that their posts come from passionate individuals.

**Collaboration and Competition**
Although there is certainly a fair amount of altruism in these people's participation, our conversations with curators gave us the impression that they are, in a way, competing with each other for attention. The one concern shared by most of the people we talked to was the issue of "tweet theft." As "Claudia" said, "I do not like that people often do not give credit in the retweets of my tweets," and when asked how this could be solved she replied, "it would be about the ability to be treated with importance, be given credibility, which would let more Twitter users know about you." Similarly, "Angela" complained that not only do others steal her tweets but they also change the timestamp she manually includes in many of her tweets to claim having scooped the news.

Being the first person to report on something is partly a matter of luck. "Angela" mentioned having her follower count increase significantly after one of her family members sent her the first photo of an event involving some people being hung from a bridge. The photo was gruesome but it spread quickly on Twitter, giving her a lot of followers. "Claudia" explained that most of her followers came from retweets that spread from friend to friend and through people searching for the relevant hashtags. Presumably, given the high volume of tweets and retweets from these curators, a search for the city hashtag would be likely to return tweets by some of these curators.

**Lack of Trust among Curators**
The second most striking issue mentioned by those we asked was the lack of trust and cooperation among curators. When asked why, one of the curators said: "I can't recommend anyone" and that "after what happened with @AnonCurator[3], I can't vouch for anybody, because I don't know who is behind that account. They are anonymous." This curator is one of the few that shows her face and has mentioned her real name publicly; the others go at lengths to maintain a high degree of anonymity. For example, "Alex" mentioned that "perhaps [the curator accounts use female names] to appear more trustworthy."

The scandal referenced by the curator was one reported on Blog del Narco [41] claiming that "military intelligence" had found evidence to suggest that @AnonCurator—one of the leading curators in the city—had been tricking citizens into "hawking" or reporting the location of the military convoys. Despite these and other suspicions, news agencies have been unable to validate whether cartels are using social media for that purpose [38].

Furthermore, even organizations like CIC are not immune to criticism; "Angela" mentioned that CIC is:

---
[3] Fictitious name

"an association that presumes not having political objectives, but obviously, there are entrepreneurial-political interests. […] not because they are in politics they are doing something wrong, not at all, that's great! But saying that it's from citizens for citizens, without political goals, I mean, don't say something you really are not."

**Sources**
Two of the people we spoke with said they had received most of the reports from their followers looking to amplify their messages, although "Claudia" acknowledged obtaining information from traditional media from time to time: "most of the information is through los tuiteros, my followers. In other cases it's the reporters on television, local news or newspapers." "Angela" on the other hand explained that she obtained most of the reports from offline sources, mostly from acquaintances or friends of friends:

> "[N]ot all the information comes in from Twitter, there's a lot of people I know, who know what I do, [people] have my phone and give me a call and they keep me informed, they are citizens, 100% citizens."

While trust regularly emerged as an issue for these curators, they often struggled to explain how they vetted their sources. When asked how to identify trusted tweets, "Claudia" simply replied "you use your instincts."

## DISCUSSION
Nationally, there have been several efforts focused on curating information. One of the most popular is called "Blog del Narco," which has been the subject of much controversy for publishing gruesome drug war images and videos that traditional media does not publish [17]. The blog went through several challenges, claiming to be attacked by hackers and targeted for censorship by the government through Google's Blogger platform [13,16] until it eventually merged with Notirex, a lesser-known news website [4].

Although traditional journalists regularly serve as curators, both on Twitter and in the more mainstream media outlets, the rise of citizen curators suggests that existing outlets are not meeting public need. Both government officials and journalists have idiosyncratically engaged on Twitter, but much of the citizen curators' success in building an audience stems from their willingness to curate information even when government agencies, journalists, and other media outlets are not.

The rise of social media as an alternate information channel has enabled the emergence of civic media curators, some of which have turned into more institutionalized civic organizations. This nascent form of civic engagement is new in Mexico and has the potential of becoming an important force when traditional institutions are weak. Yet, this practice is not without controversy. Several journalists have raised skepticism about the potential for social media to spread fear and misinformation [15]. Indeed, the fear of inaccurate information spreading has prompted government agencies to clamp down on citizen curators.

Citizen curators have been targeted both by governmental agencies and the cartels. For example, the government of Veracruz charged two people with "terrorism" for allegedly spreading misinformation on Twitter [19]; in Nuevo Laredo—one of Reynosa's neighboring cities—two bodies were found with visible signs of torture along with a placard that read: "This is going to happen to all the internet busybodies."

Yet, the popularity and prevalence of tweets related to the Drug War highlights that citizens are turning to Twitter to obtain and share information about the happenings in key cities where violence is occurring, as illustrated via our quantitative analyses.

Manuel Castells argues that insurgent communities of practice emerge in resistance to oppression when networked individualism and communalism collide [8]. These communities are constituted around a shared practice with a common set of beliefs. They are both individually motivated and committed to creating something of collective value. Citizen media curators and those contributing to popular hashtags are engaged in a shared project to inform, yet they are also driven by personal goals to gain status and attention. They are collectively engaged in resistance to both the flawed media ecosystem and the powerful role of the drug cartels; however, they are simultaneously mistrustful of each other. As an insurgent community of practice, they challenge media and governmental power by restructuring the networks through which information flows. Yet, in achieving networked power, they can both make themselves vulnerable to misinformation and also susceptible to de-anonymization, arrest, and even death.

Curators and others engaged in spreading information about the Mexican Drug War must balance between their desire to contribute to the public good, their personal interests in achieving attention or notoriety, and the vulnerabilities they face by becoming too visible or contributing too much. This tension shapes how people participate in this ecosystem.

**CONCLUSIONS**
We have presented how people use Twitter in the Mexican Drug War, an ongoing armed conflict in a country with significant Internet penetration and a highly visible presence of social media in public life. As we have described, for many Mexicans, social media has become a fluid and participatory information platform that augments and often replaces traditional news media and governmental institutions. Citizens in these localities are using social media as a resource for alerts and information dissemination.

As social media increasingly permeates society, its role in armed conflicts is poised to rise in prominence. From the Arab Spring to Occupy Wall Street, social media is one of the information platforms used in a wide range of movements. In this paper, we described how these technologies are currently used in war-like environments, and what their promises and challenges might be in the future. We illustrated the volume of activity and the number of people using social media to alert and connect with others during recurrent acute crises. We found that alerts containing location are among the most popular forms of participation. We then identified the small but influential role some users take as curators and presented their motivations and concerns.

In most crisis informatics work—and in many studies involving online communities—people are willing to trust strangers, either because the potential cost is minimal or simply because the potential gains (e.g., needed information, friendship) outweigh the costs. Yet, for those who are sharing information in an environment where doing so is risky, the potential cost of sharing is much higher. In many ways, what Mexican citizens are experiencing is the type of information ecosystem that is normally felt by criminals in an environment where they might get caught/punished. Yet, in a lawless warzone, the tables are turned.

**Implications in Social Computing Research**
This dynamic creates new challenges for designers. On the one hand, designers want to empower marginalized citizens to have a safe space in which to speak and share valuable information. On the other hand, the same tools that help oppressed people safely navigate information in a troubled society also serve as valuable tools for criminals in a lawful one. As a result, studies like ours reveal a design conundrum that raises serious questions about free speech, rights, freedoms, and protections in a digital ecosystem.

As designers grapple with this challenge, one intervention becomes apparent: there is a significant need for developing technical strategies to assess trust without revealing identity information. Most identity schemes focus on assessing whether or not someone is who they say they are. Indeed, Twitter has begun implementing "verified" identities. Verification, while valuable for certain types of interaction, is not necessarily what the curators we observed and interacted with need. While verification may be a decent proxy, all they really need to know is whether or not the information that is being presented is credible.

One way in which people seek to determine credibility of information is through reproducibility. Journalists, for example, check their sources by trying to see if other sources, disconnected from the original, have similar facts. It may be possible to simulate this in an online environment. For example, imagine that a user could see who else on Twitter posted a similar message and visually see the relationship between the posters. If information was only posted by people who were all connected to one another, it may be less reliable than if information was posted by a disparate collection of disconnected nodes. Other types of attributes of the Twitter interface have also been found to provide helpful proxies to evaluate credibility, such as profile pictures, as well as use of real names [28], or even links to external facing personal webpages of well-known individuals.

We have shown one context in which social media creates an alternate "user-generated" channel of communication that can address weaknesses in information flow. However, this new channel comes with its own credibility challenges such as issues of trust, reputation, and misinformation. Without having strong signals for interpreting the trustworthiness of the available information or the reputation of the people, viewers struggle to assess the quality of available information. This is not a new issue [18], but it becomes more salient in high stakes situations like the one described in this paper. Misinformation can have serious consequences, both for those who are consuming misinformation as well as those who are spreading it. Trust has a different valence when people fear for their safety. While more research is needed, based on our early observations of this armed conflict, we hope these insights be of value to system designers and crises responders.